\documentstyle[12pt, psfig]{article}
\newcommand{\NP}{{\em Nucl.\ Phys.\ }}
\newcommand{\PL}{{\em Phys.\ Lett.\ }}
\newcommand{\PR}{{\em Phys.\ Rev.\ }}

\newcommand{\MPL}{{\em Mod.\ Phys.\ Lett.\ }}
\newcommand{\PRL}{{\em Phys.\ Rev.\ Lett.\ }}

\newcommand{\tr}{{\rm Tr}}

\newcommand{\im}{{\rm Im}}

\newcommand{\inn}{\!\cdot\!}

\begin{document}
\pagestyle{plain}
\setcounter{page}{1}

\baselineskip16pt

\begin{titlepage}

\begin{flushright}
PUPT-1657\\
hep-th/9610250\\
\end{flushright}
\vspace{20 mm}

\begin{center}
{\huge Perturbative Dynamics of Fractional Strings on Multiply Wound D-strings
}

\vspace{5mm}

\end{center}

\vspace{10 mm}

\begin{center}
{\large Akikazu Hashimoto}

\vspace{3mm}

Joseph Henry Laboratories\\
Princeton University\\
Princeton, New Jersey 08544

\end{center}

\vspace{2cm}

\begin{abstract}
\noindent Fractional strings in the spectrum of states of open strings
attached to a multiply wound D-brane is explained.  We first describe
the fractional string states in the low-energy effective theory where
the topology of multiple winding is encoded in the gauge holonomy. The
holonomy induces twisted boundary conditions responsible for the
fractional moding of these states.  We also describe fractional
strings in world sheet formulation and compute simple scattering
amplitudes for Hawking emission/absorption. Generalization to
fractional DN-strings in a 1-brane 5-brane bound state is described.
When a 1-brane and a 5-brane wraps $Q_1$ and $Q_5$ times respectively
around a circle, the momentum of DN-strings is quantized in units of
$2 \pi/L Q_1 Q_5$.  These fractional states appear naturally in the
perturbative spectrum of the theory.
\end{abstract}

\noindent

\vspace{2cm}
\begin{flushleft}
October 1996
\end{flushleft}
\end{titlepage}
\newpage

\renewcommand{\baselinestretch}{1.1}  
\addtolength{\baselineskip}{0.5ex}

Dynamics of strings and D-branes \cite{dlp,polchinski} have been a
subject of interest recently.  Perturbatively, such a system can be
described using world sheets with Dirichlet boundary conditions
\cite{KT,GHKM,gm,decay,hashimoto96,bachas,Lifschytz:1996a}.  Upon
compactification, a new element enters the story: winding states of
D-branes and strings.  In \cite{DasMathur96,MaldaSuss96}, a
configuration where a D-brane wraps multiple times around a cycle was
introduced.  These configurations are motivated by S-duality
\cite{DasMathur96} and are needed to account for the entropy of
D-branes in the ``fat black hole limit'' \cite{MaldaSuss96}.
Excitations of these branes have been referred to in the literature as
``fractional strings.'' Their momentum is quantized in units of
$2\pi/nL$ where $n$ is the winding number of the D-brane and $L$ is
the period of the cycle.  Quantization of momenta in fractional units
can be motivated intuitively as fluctuations of a long string of
length $nL$.  They can also be understood as S-dual of excited states
of ordinary strings \cite{DasMathur96}, whose spectrum is given by
$$M = \sqrt{ (n T L)^2 + 8 \pi T N} = n T L + \frac{4 \pi N}{nL} + \ldots \ .$$
Interesting interpretation for the spectrum of fractional strings in
string field theory language was described in \cite{DMVV96}.  What is
missing is a D-brane formulation of these fractional strings.

In this note, we explain how fractional strings arises naturally in
the spectrum of states of open strings attached to a multiply wound
D-brane. Taking doubly wound D-string as a concrete example, we first
explain the fractional states in the spectrum of low energy effective
theory.  Then, we explain how the same physical spectrum follows from
the spectrum of vertex operators in the world sheet sigma model.
Using this language, we compute simple scattering amplitudes involving
these fractional strings. Finally, we discuss generalizations to
fractional states in D-branes with higher winding number and
fractional DN-strings.

\section{Multiply wound D-strings and Wilson-line Backgrounds}

We begin by describing multiply wound D-branes.  We will consider type
IIB theory and compactify the $X_1$ direction into a circle of period
$L$. To be concrete, let us focus on D-strings wound exactly twice.  A
closely related configuration carrying the same amount of
Ramond-Ramond charge is a bound state of two D-strings, each winding
once around the circle, and its spectrum includes a $U(2)$ Yang Mills
gauge fields living on the D-string world volume \cite{ed}. As was
explained in \cite{joeReview}, these two configurations are
distinguished by the presence of non-trivial gauge holonomy in the
background.  The bound state of two singly wound D-strings have the
trivial holonomy
$$U = \left( \begin{array}{cc} 1 & 0 \\ 0 & 1 \end{array} \right)$$
whereas a single long string wound twice have the non-trivial holonomy
$$U = \left( 
\begin{array}{cc} 
	0 & 1 \\
        1 & 0 \end{array} 
\right).$$
This makes perfect sense, as it means charges on one branch of the
doubly wound string returns to the other branch as it parallel
transports around the cycle once, and returns to the original branch
after parallel transporting around the cycle once more.
\begin{figure}
\centerline{\psfig{figure=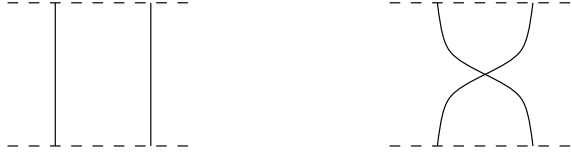}}
\caption{Schematic illustration of bound state of two singly wound D-strings and a long D-string wound twice.}
\label{figa}
\end{figure}

Gauge holonomies are due to Wilson lines and are related by
$$
U = P [e^{i \oint A}].
$$
The main point then is that multiply wound D-string can be studied
perturbatively by expanding around a vacuum with a Wilson line
background.

\section{Fractional strings in low-energy effective action}

In this section, we describe how the spectrum of fractional string
states arises in the low energy effective theory.  The low energy
effective theory on the world volume of D-string with $Q_1=2$ is the
$U(2)$ supersymmetric Yang Mills theory on $d=2$ with adjoint matter
hypermultiplets.
\begin{equation}
S = \tr \left( \left( \partial_\mu A_\nu - \partial_\nu A_\mu + [A_\mu, A_\nu] \right)^2
+ \left(\partial_\mu \Phi_M + [A_\mu, \Phi_M]\right)^2
+ [\Phi_M, \Phi_N]^2 \right)
\label{loweff}
\end{equation}
Indecies $\mu$ and $\nu$ run over world volume dimensions $\{01\}$ and
$M$ and $N$ run over transverse dimensions $\{23456789\}$.  In order
to specify the configuration where one long string is wound twice, we
set the Wilson line background to
$$
U = P[e^{i \oint A}] = \left(\begin{array}{cc} 0& 1 \\ 1 & 0 \end{array} \right).
$$
It will be convenient to work in Coulomb gauge.  In this gauge, $A_1$
will be constant along $X_1$.  Turning on a Wilson line background is
seen to be equivalent to setting the vacuum expectation value of $A_1$
to
$$
\langle A_1 \rangle = \Lambda = \frac{\pi}{2L} \left(
\begin{array}{cc}
-1 & 1 \\
 1 & -1
\end{array}
\right)
$$
Due to the presence of commutator term in the action (\ref{loweff}),
vacuum expectation value of $A_1$ will give rise to a non-standard
kinetic term.  This can fixed by invoking a well known relation which
exchanges Wilson line backgrounds with non-trivial boundary
conditions.  Consider a gauge function
$$ g(X_1) = e^{i\Lambda X_1}$$
and define $a_\mu$ and $\phi_M$ by relation
\begin{eqnarray}
A_\mu &=& (\partial_\mu g) g^{-1} + g\, a_\mu\, g^{-1}
\nonumber \\
\Phi_M & = & g\, \phi_M \, g^{-1} 
\label{redef}
\end{eqnarray}
Since $g(X)$ is not globally defined (it is not single valued under $X
\rightarrow X+L$), this is not a gauge transformation. It 
should instead be thought of as a field redefinition.

The form of the action is unchanged by the field redefinition
$$ S = \tr \left( \left
( \partial_\mu {a}_\nu - \partial_\nu {a}_\mu + [{a}_\mu, {a}_\nu] \right)^2
+ \left(\partial_\mu {\phi}_M + [{a}_\mu, {\phi}_M]\right)^2
+ [{\phi}_M, {\phi}_N]^2 \right).
$$
Vacuum expectation value of $a_\mu$ vanish, so the kinetic term of
this action is standard, and the Wilson line is trivial. In exchange,
the field $\phi_M$ acquires a non-trivial boundary condition as it
goes around the circle. 
$$
\phi_M(X_1+ L) = U^{-1}\, \phi_M(X_1)\, U.
$$

Now we can understand the spectrum and the vacuum structure in the
matter sector. Let us expand the matter hypermultiplet into components
of $U(2)$ adjoint representations
$$
\phi_M = \phi_M^0 \sigma^0
+ \phi_M^1 \sigma^1
+ \phi_M^2 \sigma^2
+ \phi_M^3 \sigma^3.
$$
In terms of these components, the boundary condition reads
$$ \phi^0(x+L) = \phi^0(x), \qquad \phi^1(x+L) =  \phi^1(x)$$
$$ \phi^2(x+L) = -\phi^2(x), \qquad \phi^3(x+L) =  -\phi^3(x),$$
that is, $\phi^0$ and $\phi^1$ are periodic and $\phi^2$ and $\phi^3$
are antiperiodic. The periodic fields are integrally moded while
anti-periodic fields are half-integrally moded. The combined spectrum
is quantized in units of $2 \pi/ n L$ which is the spectrum of
fractional strings we set out to find.

Vacuum structure in the matter sector is determined by the expectation
value of the zero modes.  Since $\phi^2$ and $\phi^3$ are
antiperiodically moded, these fields do not have zero modes.  The
$\phi^0$ simply corresponds the center of mass. The vacuum is
parameterized by the zero modes of $\phi^1$ which takes value on
$R/Z_2$.

\begin{figure}
\centerline{\psfig{figure=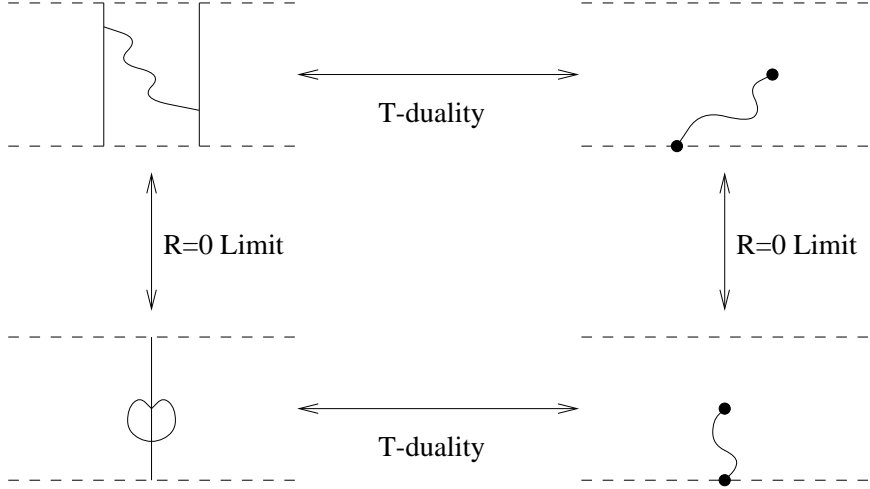}}
\caption{Stringy illustration of fractional strings and its T-dual}
\label{figb}
\end{figure}

When $\phi^1$ has a non-vanishing expectation value, $\phi^2$ and
$\phi^3$ acquire a mass and the fractional strings disappear from the
low-energy spectrum. Only $\phi^0$ and $\phi^1$ survive in the
low-energy spectrum. Physical interpretation of this configuration
becomes more clear if we diagonalize $\phi^1$ by performing a global
$U(2)$ transformation. In this basis, the vacuum expectation value of
$A_1$ becomes
\begin{equation}
A_1 = \frac{1}{L} \left( 
\begin{array}{cc}
0 & 0 \\
0 & \pi 
\end{array}\right)
\label{hol}
\end{equation}
Turning on the vacuum expectation value of $\phi^1$ then corresponds
to separating a pair of singly wound D-string.  The gauge group $U(2)$
is broken to $U(1) \times U(1)$ subgroup which corresponds to gauge
fields living on each of the branes.  The vacuum expectation value of
these $U(1)$ gauge field can be read off from the diagonal of
(\ref{hol}). These vacuum expectation values specifies a unique
superposition of fundamental winding states bound to the D-branes as
was described in \cite{ed}. This also provides a concrete physical
picture for the higgsing of the fractional states.  The fractional
states corresponds to strings stretched between the two wound
D-strings with the above $U(1)$ expectation values.  They become light
when the two D-strings coincide.  The fact that these states are
fractional can be most easily seen by T-dualizing along the $X_1$
direction.  Under T-duality, the period of the cycle is inverted and a
D-string becomes a zero brane. Since
$$4 \pi A_1 = \frac{4 \pi}{L}
\left( \begin{array}{cc}
0 & 0 \\
0 & \pi 
\end{array} \right) 
= \frac{L'}{2}
\left( \begin{array}{cc}
0 & 0 \\
0 & 1
\end{array} \right) 
$$
where $L'$ is the dual period, this configuration corresponds to that
of a two zero branes separated by half a period \cite{joeReview} (see
figure \ref{figb}).  This was anticipated in \cite{mathur96} but here
we derived it directly using the low energy effective description of
the D-brane.  The open string stretching from one zero brane to the
other must then wrap around the cycle by half integral multiple of the
period.  T-dualizing back to a D-string picture, we find that the
momentum of these states are half integrally moded, which is precisely
what we expect for the fractional strings.

\section{World sheet description of multiply wound D-strings}

In this section, we extend the result of the previous section to full
string theory.  To this end, it suffices to consider an open string
non-linear sigma model in a Wilson line background. The spectrum of
fractional strings can be read off from the spectrum of marginal
operators of this sigma model. Once the vertex operators are
constructed, correlation functions can be computed in a
straightforward manner.

The basic structure of non-linear sigma models in a Wilson line
background is quite simple.  One simply adds a coupling to background
gauge field in the Polyakov path integral.
\begin{equation}
\langle \ldots \rangle = 
\tr [P\,\int DX ( \ldots )\  e^{-S + i \oint  A_1 \partial_t X_1}  ]
\label{CFT}
\end{equation}
Processes involving fractional strings are sometimes easier to
visualize in the T-dual description where the string winds by a
fractional period. The T-dual of the above path-integral expression is
\begin{equation}
\langle \ldots \rangle = 
\tr [P\,\int DX ( \ldots )\  e^{-S + i \oint  A_1 \partial_n X_1}  ]
\label{CFT2}
\end{equation}
Care is needed to path-order the exponential as the gauge field $A$ is
non-abelian. This only affects the vertex operators sitting at the
boundary of the world sheet. Of course, fractional strings are such
operators.  To be concrete, let us consider computing an amplitude
involving two boundary vertex operators and arbitrary number of bulk
vertex operators. Due to path ordering, the correlation function takes
the form
$$
\langle V_1(\sigma_1) V_2(\sigma_2) (\ldots) \rangle
=   \int DX\,  (\ldots) \,
\tr[ V_1(\sigma_1) U(\sigma_1,\sigma_2) V_2(\sigma_2) U(\sigma_2,\sigma_1)\,
] \, e^{-S} 
$$
where 
$$U(\sigma_1,\sigma_2) = P[ e^{i \int_{\sigma_2}^{\sigma_1} A_1
\partial_n X_1}]$$
The only non-trivial effect of $U(\sigma_1,\sigma_2)$ is its action on
the Chan-Paton factor of $V(\sigma)$. Since we are interested in the
physics of fractional strings, let us set the Chan-Patton factors $T_1
= T_2 = \sigma^3$ and examine the expression
$$ \tr[T_1 U(\sigma_1,\sigma_2) T_2 U(\sigma_2,\sigma_1)] $$ 
more closely.  Just as in the previous section, it is convenient to
work in Coulomb gauge where $A_1 = \Lambda = {\rm constant}$.
Using the fact that 
$$\sigma^3 \Lambda \sigma^3 = \frac{\pi}{2L}\left(
\begin{array}{cc}
-1 & -1 \\
-1 & -1 
\end{array}\right)$$
one can easily show that
$$\tr[
T_1 U(\sigma_1,\sigma_2) T_2 U(\sigma_2,\sigma_1) ]
= e^{-i \int_{\sigma_2}^{\sigma_1} \frac{\pi}{L} \partial_n X}+ 
e^{-i \int_{\sigma_1}^{\sigma_2} \frac{\pi}{L} \partial_n X}.
$$ 
In the electrostatic language, this corresponds to placing an electric
dipole with its moment normal to the boundary of the world sheet,
which has precisely the effect of turning on an electrostatic
potential difference between the two components of the world sheet
boundary marked by $\sigma_1$ and $\sigma_2$.  This is precisely the
expected world sheet description of string states which stretches
between two 0-branes separated by distance $L'/2 = (2\pi)^2/L$.

\begin{figure}
\centerline{\psfig{figure=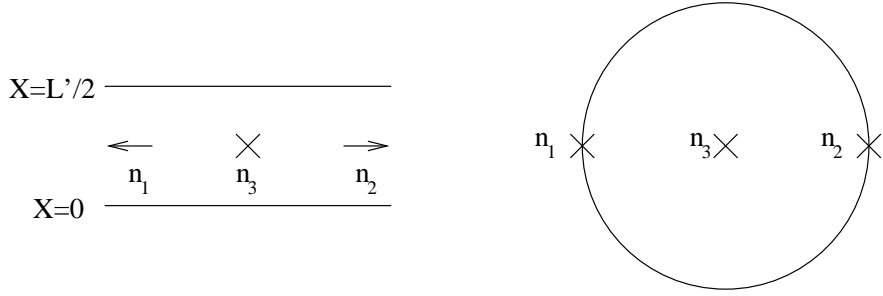}}
\caption{String world sheet amplitude for a process where a wound
closed string with momentum $q$ and polarization $\varepsilon_{MN}$ is
absorbed by one of the zero branes, creating a pair of fractionally
wound open string states with winding numbers $p_1$ and $p_2$.}
\label{figc}
\end{figure}

Now let us examine the stringy dynamics of these fractional winding
states. To this end we must construct vertex operators for these
states and compute their correlation function.  An essential
ingredient in such a vertex operator is the piece which encodes the
fact that the string is stretched between separated D-branes. This can
be determined from the singularity structure in the electric field at
the boundary changing point as follows. First consider a strip $0 \le
\im[\sigma] \le \pi$ with boundary condition $X(\sigma)=0$ at
$\im[\sigma]=0$ and $X(\sigma)=V$ at $\im[\sigma]=\pi$.  The
background $X(\sigma)$ which satisfies this boundary condition is
$$X(\sigma) = \frac{V}{\pi} \im[\sigma].$$
Now map this strip to a half-plane using the map $z =
e^{\sigma}$. This maps the boundary changing points to $\sigma=0$ and
$\sigma=\infty$. In the $z$-plane,
$$X(z,\bar{z}) = \frac{V}{2 \pi i} (\ln (z) - \ln (\bar{z}))$$
so near $z=0$,
$$\partial X(z) = \frac{V}{2 \pi i} \frac{1}{z}$$
and
$$\bar{\partial} \bar{X} (\bar{z}) = -\frac{V}{2 \pi i} \frac{1}{z}$$
This is precisely the structure of operator product expansion
$$\partial X(z) e^{iqX }e^{- iq\bar{X}}(0) = 
\frac{2iq}{z} e^{iqX}e^{-iq\bar{X}}(0)$$
$$\bar{\partial} \bar{X}(\bar{z}) e^{iqX}e^{ - iq\bar{X}}(0) 
= -\frac{2iq}{\bar{z}} e^{iqX}e^{-iq\bar{X}}(0)$$
for 
$$ q = -\frac{V}{4 \pi} = -\frac{L'}{8\pi} = \frac{\pi}{L}.$$
On the boundary, $X = - \bar{X}$, so 
$$ e^{iqX} e^{-iq\bar{X}} = e^{ i\, 2q X}.$$
which effectively shifts the winding number by $q$.  By now the basic
construction of the vertex operator for the fractional string state is
clear. World sheet supersymmetry constrains the form of the vertex
operator to take the conventional form
$$ V(z) = (\partial X^\mu + i (2k \inn \psi) \psi^\mu) e^{i\, 2k X}(z)$$
but the boundary condition shifts the winding number
$$ n =\frac{4 \pi}{L'}k_1$$
to take on half-integral values. So we see how fractional string
spectrum arises from the world sheet considerations.  These vertex
operators can be used to compute amplitudes involving fractional
strings. Perhaps the simplest such amplitude is a process where a
macroscopic string \cite{dh} in a ground state of winding number $n_3$
with momentum $\vec{q}$ is absorbed by one of the zero branes,
creating a pair of fractionally wound open string states with winding
numbers $n_1$ and $n_2$. (see figure \ref{figc}.)
The amplitude takes the general form
$$ A = \int \frac{dz_1\, dz_2 \, d^2 z_3}{V_{CKG}}
\langle V_1(z_1) V_(z_2) V_3(z_3,\bar{z}_3) \rangle$$
where
\begin{eqnarray}
V_1(z_1) & = &  \xi_M^1 (\partial X^M + i (2p_1 \inn \psi) \psi^M) 
e^{i\, 2p_1 X}(z_1) \nonumber \\
V_2(z_2) & = &  \xi_M^2 (\partial X^M + i (2p_2 \inn \psi) \psi^M) 
e^{i\, 2p_2 X}(z_2) \nonumber
\end{eqnarray}
for fractional strings and
$$V_3(z_3,\bar{z}_3) = \varepsilon_{M N}\, 
(e^{-\phi} \psi^M e^{i q_L X}(z_3)) \, 
(e^{-\bar{\phi}} \bar{\psi}^N e^{i q_R \bar{X}}(\bar{z}_3)). $$
for the macroscopic string with
\begin{eqnarray}
p_1 & = & (p_1^0, \frac{L'}{4 \pi} n_1) \nonumber \\
p_2 & = & (p_2^0, \frac{L'}{4 \pi} n_2) \nonumber \\
q_L & = & (q^0, \frac{2\pi}{L'} m_3 + \frac{L'}{4 \pi} n_3, \vec{q}) \nonumber \\
q_R & = & (q^0, \frac{2\pi}{L'} m_3 - \frac{L'}{4 \pi} n_3, \vec{q}) \nonumber
\end{eqnarray}
and $m_3=0$. This is precisely of the form of the amplitude computed
in \cite{decay} so we simply quote the result
$$ A = \frac{\Gamma(-2t)}{\Gamma(1-t)^2} t^2 
\left( \xi^1 \inn \varepsilon \inn \xi^2 + 
\xi^2 \inn \varepsilon \inn \xi^1 \right)
$$
where 
$$t = - 2p_1 \inn p_2$$
with the requirement that $n_1$ and $n_2$ take on half-integer values.
Of course, by T-duality, this amplitude is exactly equivalent to the
amplitude for a process where a graviton of momentum $q$ is absorbed
by a doubly wound D-string creating a pair of fractional momentum open
strings with momentum $p_1$ and $p_2$.

\section{Generalizations to multiply wound DD and DN strings}

Until this point we have focused on fractional excitations of D-string
wound exactly twice.  The basic structure underlying the fractional
states are the $U(2)$ holonomy matrix
$$U = \left(
\begin{array}{cc}
0 & 1 \\
1 & 0 
\end{array} \right)$$
and its action on the $U(2)$ generators
$$\tr[T U^{-1} T U] = \pm 1.$$
The sign of $\tr[T U^{-1} T U]$ induced the twisting of the boundary
condition which gives rise to fractionally moded states in the
low-energy dynamics, and was ultimately responsible for the placement
of 0-branes separated by half a period in the T-dual description.

This structure has a natural generalization to configurations with
higher winding numbers.  The low-energy effective theory for the
$n$-tuply wound D-string configuration is a $U(n)$ gauge theory with
the holonomy
\begin{equation}
U = \left(\begin{array}{cccc}
0 & 1 &        & \\
  & 0 & \ddots & \\
  &   & 0      & 1 \\
1 &   &        & 0
\end{array}
\right).
\label{genhol}
\end{equation}
Twisting of the boundary is induced by the phase factor given by the
eigenvalues of the matrix
$$
M_{ab} = \tr[ T_a U^{-1} T_b U].
$$
It turns out that for all $n$, the spectrum of eigenvalues is
precisely the $n$-fold degenerate spectrum of
\begin{equation}
\left\{ 1,\ e^{2\pi i /n},\ e^{4\pi i/n},\ \ldots,\ e^{2(n-1)\pi i/n} \right\}
\label{spectrum}
\end{equation}
which implies a combined spectrum quantized in units of $2 \pi /
nL$. 

To see that the spectrum of $M_{ab}$ is indeed of the form
(\ref{spectrum}), simply note that adjoint representation of $U(n)$ is
a tensor product $n \otimes \bar{n}$. Natural basis for the
fundamental representation of $U(n)$ in this context the eigenvectors
$v_a$ of the holonomy matrix $U$ with eigenvalues $\lambda_a = \exp(2
\pi i a/n)$. An adjoint element $v_a \otimes \bar{v}_b$ transforms
under $U$ like
$$ U^{-1} (v_a \otimes \bar{v}_b) U = e^{2 \pi i (a-b)/n}(v_a \otimes \bar{v}_b)
$$
where $a$ and $b$ takes integer values between $0$ and $n-1$.  Now,
$(a-b) {\rm \ mod\ } n$ is an $n$-fold degenerate set of integers
ranging from 0 to $n-1$. Perhaps an example would suffice to make this
point. For $n=3$,  $((a-b) {\rm \ mod\ } n)$ takes on values
$$
\begin{tabular}{|c|c|c|c|} \hline
  & b=0 & b=1 & b=2 \\ \hline
a=0 & 0 & 2 & 1 \\ \hline
a=1 & 1 & 0 & 2 \\ \hline
a=2 & 2 & 1 & 0 \\ \hline
\end{tabular}
$$
as is expected.

We can generalize these ideas further to fractional states in the
Dirichlet-Neumann (DN) sector.  DN strings appear when branes of
different dimensionality, say a 1-brane and a 5-brane, are
simultaneously present.  The 1-brane 5-brane system compactified on
$T^5$ is by now familiar as a model of extremal black hole with a
regular horizon \cite{cm,ghas}.  When the 5-brane wraps around $T^5$,
it is possible in principle to turn on non-trivial holonomies for all
cycles.  Here, we will restrict our attention to the simple case where
the only non-trivial holonomy is along the direction parallel to the
1-brane.  This is essentially the case considered in
\cite{MaldaSuss96}.

Now consider a 1-brane wrapped $Q_1$ times and a 5-brane wrapped $Q_5$
times around the cycle in $X_1$ direction. This corresponds to turning
on a holonomy matrix $U_{(1)}$ of $U(Q_1)$ and $U_{(5)}$ of $U(Q_5)$
of the type (\ref{genhol}).

The DN-strings carries a Chan-Patton factor $T_{(1,5)}$ which
transforms as $Q_1 \otimes \bar{Q}_5$ and computing the extent of
twist in the boundary condition reduces to computing the eigenvalue of
$$M_{ab} = \tr[(T_{(1,5)})^{\dagger} U_{(1)} T_{(1,5)} U_{(5)}^{-1}].$$
Just as in the previous case, it is convenient to construct the basis
$T_{(1,5)}$ using the tensor product of $v_a$ and $\bar{w}_b$ where
$v_a$ is an eigenvector of $U_{(1)}$ with eigenvalues $\exp(2 \pi i
a/Q_1)$ and $w_b$ is an eigenvector of $U_{(5)}$ with eigenvalues
$\exp(2 \pi i b/Q_5)$. Then
$$ U_{(1)} (v_a \otimes w_b) U_{(5)}^{-1} 
= e^{2 \pi i (a Q_5 - b Q_1)/Q_1 Q_5} (v_a \otimes w_b) $$
Just as in the previous case, let us make a table of $((a Q_5 - b Q_1)
{\rm \ mod \ } Q_1 Q_5)$ for $Q_1$ = 2 and $Q_5=3$.
$$
\begin{tabular}{|c|c|c|c|} \hline
    & b=0 & b=1 & b=2 \\ \hline
a=0 & 0   & 4   & 2 \\ \hline
a=1 & 3   & 1   & 5 \\ \hline
\end{tabular}
$$
We see that $((a Q_5 - b Q_1) {\rm \ mod \ } Q_1 Q_5)$ takes on
integer values between 0 and $Q_1 Q_5 - 1$ with no degeneracy. This
will be true in general for $Q_1$ and $Q_5$ relatively prime, implying
that the spectrum of these states will be quantized in units of $1/Q_1
Q_5$ which is precisely what was anticipated in
\cite{MaldaSuss96}. (In general, the spectrum will be quantized in
units of $N/Q_1 Q_5$ with degeneracy $N$ where $N$ is the greatest
common divisor of $Q_1$ and $Q_5$.)  Perturbative dynamics of
fractional DN-strings can then be computed along the lines of
\cite{hashimoto96}.

\section{Conclusions}

In this article we considered the excitations on D-branes wrapped
multiple times by turning on a background with a non-trivial gauge
holonomy.  We find a spectrum of states quantized in fractional units
of momentum along the compactified direction. These are the so-called
``fractional strings'' whose existence have been postulated previously
on the grounds of entropy counting and S-duality. We have provided an
explicit check on this postulate by constructing these states in the
D-brane formulation.

In the low-energy effective theory, gauge holonomies have the effect
of twisting the boundary condition.  Fractional states arises
naturally as a result of twisted boundary conditions.

Same basic conclusion can be drawn by considering the full open-string
theory in the world sheet formulation.  Here, it was convenient to
consider the T-dual of multiply wound D-string which corresponded to
$n$ D0-branes equally spaced along the circle of compactification. The
fractional momentum states have a natural interpretation in the T-dual
picture as strings stretching between these D0-branes which winds
around the period of compactification in fractional units.  Vertex
operators for these twisted strings were constructed with which we
computed the Hawking emission/absorption amplitude.

Finally, we described how the same idea can be used to explain the
fractional spectrum of DN-strings.  For example, it is possible to
show that the momentum of DN-strings in a 1-brane 5-brane system with
winding numbers $Q_1$ and $Q_5$ are quantized in units of $1/Q_1
Q_5$. This has been anticipated in \cite{MaldaSuss96} but it is
pleasant to find an explicit D-brane construction.

Recent exciting work on string dualities has predicted the existence
of exactly one threshold bound state \cite{senmar} in the quantum
mechanics of these D-branes \cite{DanFerrSund,PouliotKabat,DKPS96}. By
construction, our discussion of fractional strings is perturbative,
and at leading order captures only the semi-classical aspects of
D-brane dynamics.  Presumably, the full quantum dynamics of D-branes
such as the existence of these bound states is encoded in the full
re-summation of scattering amplitudes, but it is not clear how this is
done in detail.  Perhaps something can be learned by examining the
large order growth \cite{Shenker:1990} of the perturbative
expansion. It would be very interesting to understand how perturbation
theory and D-brane quantum mechanics fit together in the full
treatment of non-perturbative string dynamics. In particular, we have
shown in this article that the semi-classical configuration of a
D-string wound $n$ times on period $L$ is T-dual to D0-branes equally
spaced along the dual period $L'$.  Recently, it was argued on the
grounds of S-duality that the threshold bound state approximates this
semi-classical configuration in the limit where period $L'$ is small
\cite{mathur96}.  Something interesting appears to be happening when
$L'$ is taken small enough to squeeze the 0-brane bound state wave
function to a size smaller than its natural scale.  It would be very
interesting to understand this phenomenon in detail by studying the
dynamics of D0-branes along the lines of
\cite{DanFerrSund,PouliotKabat,DKPS96} but on a compactified space.
We leave these fascinating questions for future investigations.

\section*{Acknowledgments}

We thank Rajesh Gopakumar, David Gross, Igor Klebanov, Barek Kol, Juan
Maldacena, Samir Mathur, Marco Moriconi, Vipul Periwal, Arvind
Rajaraman, Sanjaye Ramgoolam, and Washington Taylor for illuminating
discussions.  This work was supported in part by DOE grant
DE-FG02-91ER40671, the NSF Presidential Young Investigator Award
PHY-9157482, and the James S. McDonnell Foundation grant No. 91-48.

\addtolength{\baselineskip}{-0.5ex}

\end{document}